\documentclass[12pt,preprint]{emulateapj}
\usepackage{amsmath}
\usepackage{amssymb}
\usepackage{graphicx}
\usepackage{subfigure}

\def\msun{M_\odot}
\def\mbh{M_{\rm{BH}}}

\def\<{\,\langle\langle}
\def\>{\,\rangle\rangle}

\newcommand{\mbulge}{\ensuremath{M_\mathrm{BH}-L_{\mathrm{bulge}}}}
\newcommand{\mgal}{\ensuremath{M_{\mathrm{BH}}-M_{\mathrm{bulge}}}}
\newcommand{\msigma}{\ensuremath{M_{\mathrm{BH}}-\sigmastar}}
\newcommand{\sigmastar}{\ensuremath{\sigma_{\ast}}}

\begin{document}

\title{Black Hole Mass and Bulge Luminosity for Low-mass Black Holes}

\author{ Yan-Fei Jiang \altaffilmark{1}, 
	Jenny E. Greene  \altaffilmark{1} and
	Luis C. Ho  \altaffilmark{2}
	 }

\altaffiltext{1}{Department of Astrophysical Sciences, Princeton
University, Princeton, NJ 08544, USA} 
\altaffiltext{2}{The Observatories of the Carnegie Institution for Science, 813 Santa Barbara 
Street, Pasadena, CA 91101, USA}

\begin{abstract}

  We study the scaling between bulge magnitude and central black hole
  (BH) mass in galaxies with virial BH masses $\lesssim 10^6\msun$.
  Based on careful image decomposition of a snapshot \emph{Hubble
    Space Telescope} $I$-band survey, we found that these BHs are
  found predominantly in galaxies with pseudobulges.  Here we show
  that the \mbulge\ relation for the pseudobulges at low mass is
  significantly different from classical bulges with BH masses $\geq
  10^7\msun$.  Specfically, bulges span a much wider range of bulge
  luminosity, and on average the luminosity is larger, at fixed
  $M_{\rm BH}$.  The trend holds both for the active galaxies from
  Bentz et al. and the inactive sample of G\"ultekin et al. and cannot
  be explained by differences in stellar populations, as it persists
  when we use dynamical bulge masses.  Put another way, the ratio between bulge and BH mass
  is much larger than $\sim 1000$ for our sample. This is consistent with 
  recent suggestions that $M_{\rm BH}$ does not scale with the
  pseudobulge luminosity.  The low-mass
  scaling relations appear to flatten, consistent with predictions
  from Volonteri \& Natarajan for massive seed BHs.
\end{abstract}


\section{Introduction}
\label{sec:intro}

For massive elliptical galaxies and galaxies with classical bulges,
black hole (BH) masses have been found to correlate with various
properties of bulge, such as bulge mass
\citep[e.g.,][]{Magorrianetal1998}, bulge luminosity (the \mbulge\
relation, see \citealt{MarconiHunt2003}), and the velocity dispersion
of stars in the bulge (the \msigma\ relation, see
\citealt{Tremaineetal2002}).  This has led to theoretical work
exploring the importance of AGN feedback on galaxy evolution and the
creation of the scaling relations
\citep[e.g.,][]{CiottiOstriker2007}.
However, we still do not understand the origin of the scaling
relations.  Observationally, they are best-measured
using nearby elliptical galaxies with central supermassive BHs having
masses $\gtrsim10^7\msun$.  It is still unclear what the scaling
relations look like for BHs with masses $\sim 10^5-10^6\msun$ in
late-type galaxies. Some simulations
\citep[e.g.,][]{VolonteriNatarajan2009} show that the \msigma\
relation can result from the the hierarchical merging of massive dark matter
halos \citep[also see][]{Peng2007}.  In that picture, and assuming
massive BH seeds ($\sim10^5\msun$), the \msigma\ relation would flatten
at the low-mass end. Measurement of the scaling relations of
low-mass BHs will directly constrain models of BH seed formation and help
elucidate the most important factors in establishing the observed
scaling relations.

Based on \emph{Hubble Space Telescope} (\emph{HST})
observations of the sample of $18$ low-mass BHs selected from the first data
release (DR1) of the Sloan Digital Sky Survey (SDSS;
\citealt{GreeneHo2004}), \cite{Greenetal2008} study the
\mbulge\ relation at low mass and find a different scaling, with BHs
living in larger bulges than expected from the high-mass relations.
However, their sample is small and the result is quite
uncertain. \cite{Greenetal2007} present 174 low-mass BHs from DR4 of
the SDSS, giving us the opportunity to study the scaling relations
with an order of magnitude more objects. Their \msigma\ relation has
been examined by \cite{Xiaoetal2011}. We have done
detailed image decompositions of these galaxies based on \emph{ HST}
imaging \citep[][hereafter Paper I]{Jiangetal2011} and the
\mbulge\ relation will be studied here.

Low-mass BHs are found in different host galaxy types than more
massive BHs.  Most of their host galaxies actually contain
pseudobulges.  Pseudobulges have quite different properties from those
of violently and rapidly formed classical bulges. For example,
pseudobulges have flatter shapes than those of classical bulges (e.g.,
S{\'e}rsic index $n<2$).  They have smaller velocity dispersions at
fixed luminosity in the Faber-Jackson relation
\citep[][]{FaberJackson1976} compared with classical bulges (e.g.,
\citealt{KormendyIllingworth1983,KormendyKennicutt2004}; Paper I).
Pseudobulges are believed to form via secular processes in disk
galaxies (see the review by \citealt{KormendyKennicutt2004}).
Considering their differing formation mechanisms and properties, we
may expect differing BH scaling relations as well, which is observed with
small samples with dynamical BH masses (e.g.,
\citealt{Hu2008,Hu2009,Greenetal2010}) and indirect analysis (e.g.,
\citealt{GadottiKauffmann2009}).  As the bulges hosting the low-mass
BHs presented here are likely dominated by pseudobulges (Paper I),
studying the properties of these galaxies will also help us understand
the growth and evolution of pseudobulges.

\section{The Sample and Data Analysis}

The sample and observations are described in \cite{Greenetal2007} and
Paper I.  Here we briefly describe the key properties of the sample
for completeness. The 174 galaxies are selected from DR4 of the SDSS
(\citealt{Greenetal2007}) to have virial BH masses $<2\times10^6\msun$.  
They have a median redshift of 0.085 and a median $M_{\rm BH}$
of $1.2\times10^6\msun$.

The host galaxy structures and luminosities are derived from a
snapshot survey with Wide Field Planetary Camera 2 (WFPC2) on
\emph{HST} in Cycle 16.  With the Planetary Camera (PC), we achieve a
resolution of $0\farcs046$ per pixel and a field of view of
$36\farcs8\times36\farcs8$.  The observations are done with the F814W
filter (mean wavelength of $8269$\AA), which we refer to as the
$I$-band in the following. For each object we obtain a short (30
sec) exposure in case of saturation, followed by two dithered
$\sim 600$ sec exposures. Images are reduced following standard
procedures, with extra care taken to remove cosmic ray trails (see
Paper I for details).

\subsection{Black Hole Mass}

For massive BHs in the nearby universe, the most reliable BH masses
are based on dynamical tracers such as stars
\citep[e.g.,][]{Gebhardt2004}, gas \citep[e.g.,][]{Barthetal2001}, or
water masers \citep[e.g.,][]{Herrnsteinetal2005,Kuoetal2011}.
However, current instrumentation lacks the spatial resolution to
obtain dynamical measurements for such low-mass BHs outside the Local
Group.  Thus, we use virial BH masses measured from active galactic
nucleus (AGN) emission lines.  Reverberation mapping
\citep[][]{BlandfordMckee1982} provides sizes for the broad-line
regions (BLRs) of a few dozen AGNs, from which a relation between BLR
size and AGN luminosity is calibrated \citep[e.g.,][]{Kaspietal2000,
  Bentzetal2006,Bentzetal2009b}.  Then the virial mass is simply
$\mbh=fR(\Delta v)^2/G$, where $f$ is a dimensionless factor that
accounts for the unknown geometry of the BLR, $R$ is the size of BLR and $\Delta
v$ is some measure of the width of the broad-line width such as full width
at half maximum (FWHM).  In practice, the width and luminosity of the
broad H$\alpha$ emission line \citep{Xiaoetal2011} are used to
estimate the virial BH masses
\citep[e.g.,][]{GreeneHo2006}.

Despite the large uncertainty in virial masses, it has been
demonstrated that the virial masses of SMBHs are in good agreement
with masses determined based on the \msigma\ relation (e.g.,
\citealt{Gebhardtetal2000,Ferrareseetal2001,Nelsonetal2004,GreeneHo2009}).
The uncertainty in the virial masses is thought to be a factor of $\sim 3$ (e.g.,
\citealt{GreeneHo2006,Collinetal2006,Shenetal2008,Wooetal2010}).  We
adopt a single value of $f=0.75$, which is intended to represent an
ensemble average over our sample and to be consistent with other
people's work (e.g., \citealt{Kaspietal2000,Xiaoetal2011}).

\subsection{GALFIT and Bulge Luminosity}

We use detailed image decomposition to isolate the bulge component in
the $I$-band images. Following \cite{Greenetal2008}, we perform full
two-dimensional fitting with GALFIT (e.g.,
\citealt{Pengetal2002,Pengetal2010}).  Details about the fitting and the
best-fitting models for our sample are given in Paper I. Here we
briefly describe the basic elements.

A point-spread function (PSF) model from TinyTim \citep[][]{Krist1995}
is used to model the AGN component.  For other components of the host
galaxies, we use a generalized S\'ersic \citep{sersic1968} model to fit the
light.  We fix the S{\'e}rsic index $n$ to be $0.5,1,2,3,4$ so that
GALFIT will not converge on an unreasonably high $n$ value. If necessary, 
a bar is also included as an $n=0.5$ S\'ersic component.  The sky is
fitted using a variety of methods, in order to estimate parameter
uncertainties due to the sky level.  Also, we use alternate PSF models
to bracket the uncertainties due to errors in the PSF model.  The
values of the bulge luminosities and uncertainties for our sample are
given in Table 2 of Paper I.

\subsection{Properties of our sample from \cite{Jiangetal2011}}
\label{sec:PaperI}

In Paper 1 we present evidence that the majority of the disk galaxies
in our sample likely contain pseudobulges, as summarized here.  Of the
galaxies with reliable S{\'e}rsic index $n$ measurements, $70\%$ have
$n<2$. Extended disks are found in $92\%$ of galaxies, while bar
structures are identified in $40\%$ of the sample. About half of the
galaxies have bulge-to-total ratio $B/T$ smaller than $0.2$ and $74\%$
have $B/T<0.4$. All of our disk galaxies satisfy the relations for
pseudobulges given by \cite{Gadotti2009}.  Based on these properties,
and their fundamental plane scalings, we argue that the disk galaxies
predominantly have pseudobulges, and that the galaxies without disks
are bright spheroidal, rather than elliptical, galaxies
\citep{Kormendyetal2009}.  Most striking is the Faber-Jackson
relation.  At a fixed bulge magnitude, our galaxies have smaller
velocity dispersions on average.

\begin{table}
\begin{center}
\caption{Fitting result}
\label{fittingresult}
\scriptsize
\begin{tabular}{ccccc}
\hline\hline
\multicolumn{5}{c}{Fitting of the \mbulge\ relation}\\ 
\hline
Sample					& 	$\alpha$	& $\beta$	 &    $\epsilon_0$	&    $M_0$	\\
\hline
Bentz et al. 				& 	$-0.28\pm 0.03$ & $8.30\pm0.05$	&	0.22			&   $-22.45$	\\
Bentz et al. $+$ our sample 	&	$-0.43\pm 0.03$ & $6.65\pm0.05$	&	0.41			&  $-19.79$	\\
G\"ultekin et al.				&	$-0.46\pm 0.05$ & $8.30\pm0.07$	&	0.38			&  $-22.55$	\\
G\"ultekin et al. $+$ our sample	&	$-0.52\pm 0.03$ & $6.41\pm 0.05$	&	0.48			&  $-19.84$	\\
G\"ultekin et al. $+$ Kormendy et al. & $-0.46\pm 0.05$ & $8.08\pm0.08$  &	0.50			& $-22.15$	\\
\hline\hline
\multicolumn{5}{c}{Fitting for $M_{\rm BH}-M_{\rm bulge}$ relation}\\ 
\hline
Sample					& 	$\alpha$	& $\beta$	 &    $\epsilon_0$	&    $M_{\text{dyn},0}$	\\
\hline
G\"ultekin et al. 				& 	$1.04\pm 0.16$ & $8.39\pm0.45$	&	0.45			&   $11.04$	\\
G\"ultekin et al. $+$ our sample	&	$1.37\pm 0.09$ & $6.35\pm 0.10$	&	0.42			&  $9.63$	\\
\hline
\multicolumn{5}{l}{\scriptsize Notes: We fit equation \ref{fittingrelation} for 
the \mbulge\ relation and a similar formula}\\
\multicolumn{5}{l}{\scriptsize  for the \mgal\ 
relation. Here $\epsilon_0$ is the intrinsic scatter.} \\
\end{tabular}
\end{center}

\end{table}%

\begin{figure*}
\centering
\includegraphics[width=0.45\hsize]{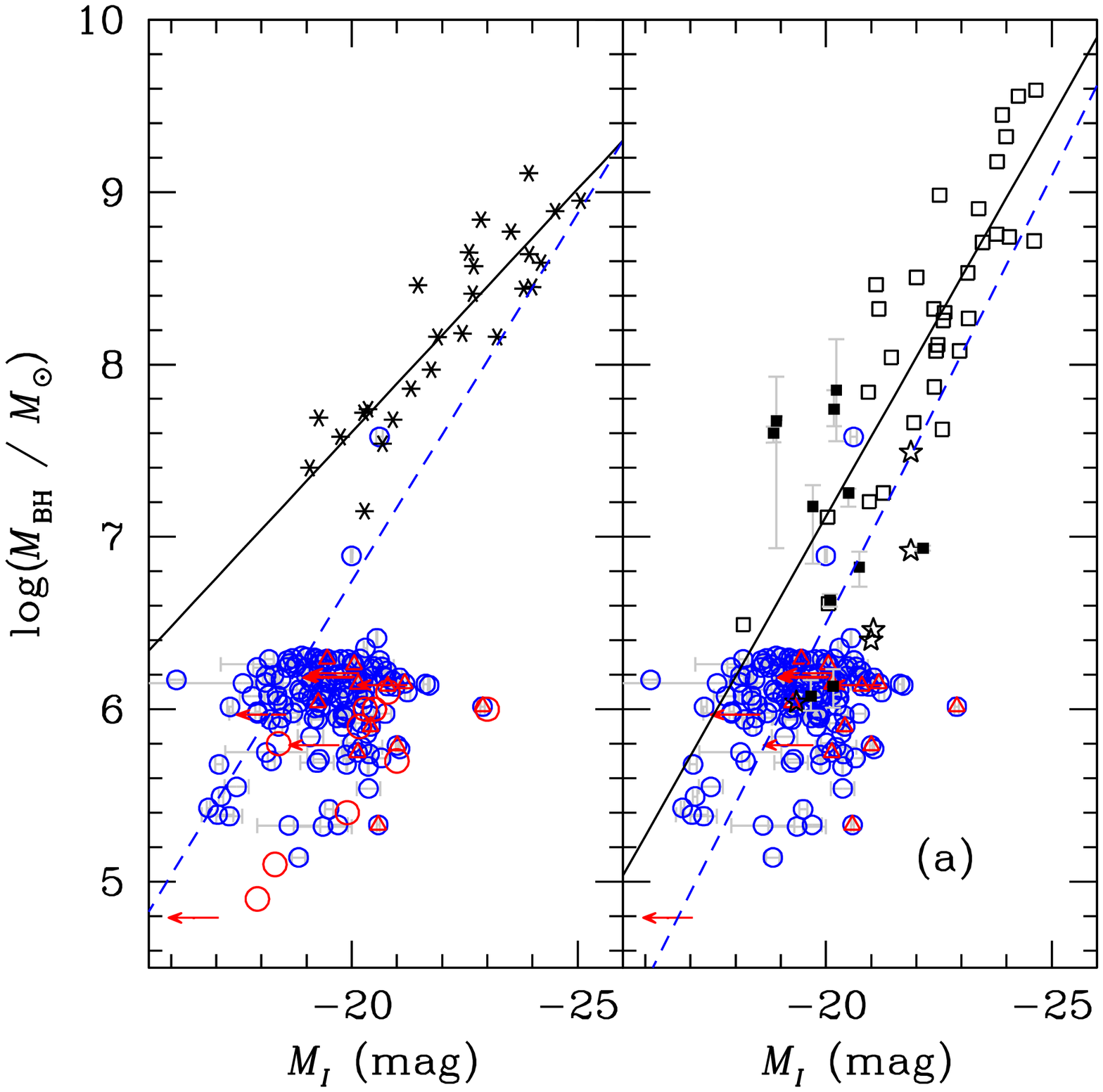}
\hspace{-1.8cm}
\includegraphics[width=0.45\hsize]{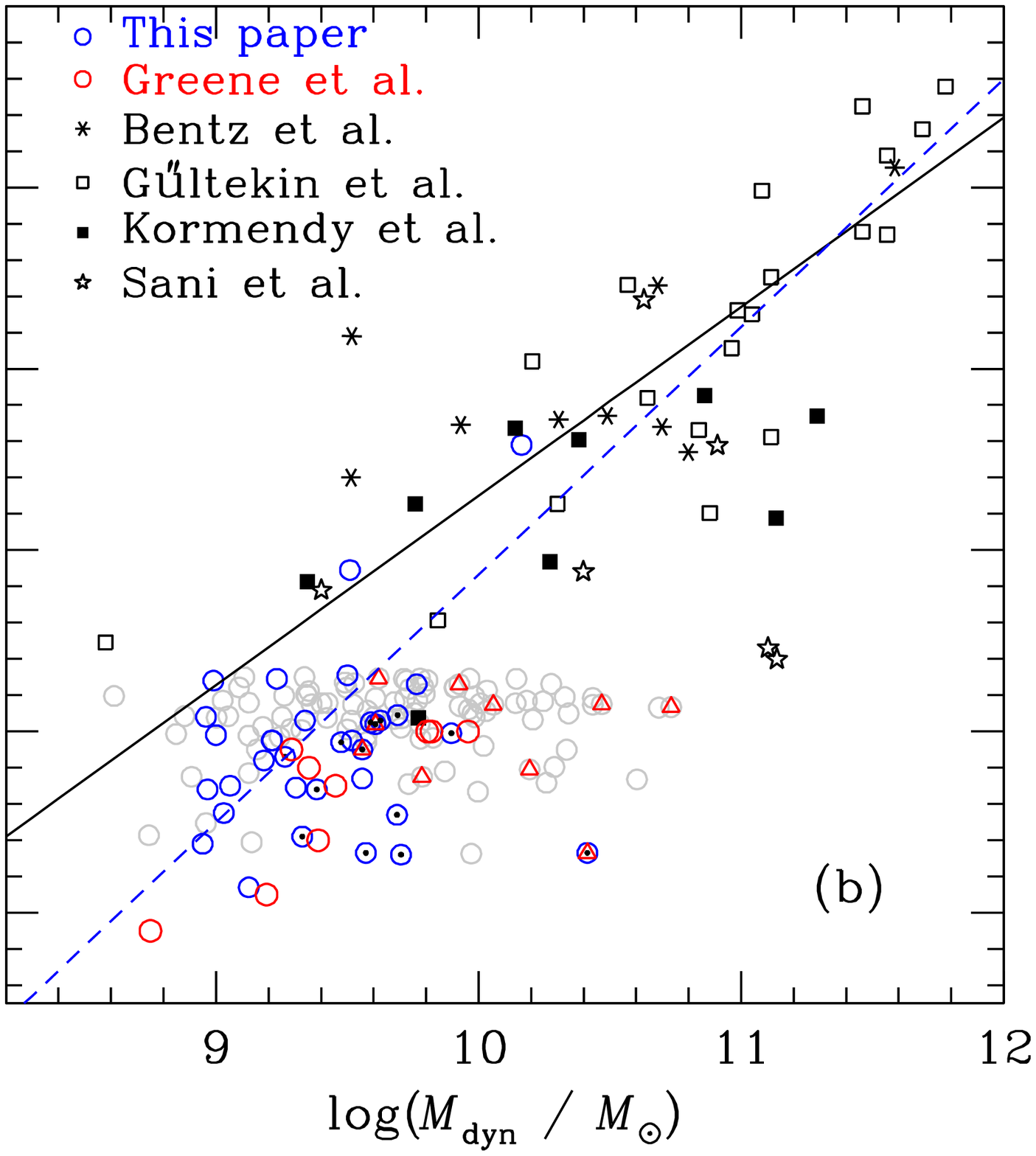}
\caption{{\bf (a)}: The relation between the BH masses $\mbh$ and $I-$band absolute 
magnitude $M_I$ of the bulge.  Our
sample is labeled with open blue circles. Galaxies without an
extended disk are labeled with red triangles. Upper 
limits of the bulges for bulgeless galaxies are plotted with 
arrows. Solid lines are the best fitting relation for active galaxies 
from \cite{Bentzetal2009a} and inactive galaxies 
from \cite{Gutekinetal2009}, respectively, in the two sub-panels. 
Dashed blue lines are the best-fitting relations when our sample is included. 
All comparison samples are 
shifted to the $I$-band appropriately. {\bf (b)}: 
The relation between BH mass and dynamical bulge mass. Grey open
circles are objects from our sample with velocity dispersions measured
from [\ion{S}{2}] \citep[][]{Greenetal2007}. Open blue circles with
black dot at the center are galaxies with little disk contamination
for the stellar velocity dispersion measurements.  We show the
best-fit relation of the G\"ultekin et al. sample using dynamical
bulge masses from \cite{HaringRix2004} (solid line) and the best fit
including the subsample of galaxies for which we have stellar velocity
dispersions (blue line) . } 
\label{MBH}
\end{figure*}

\section{Results}

We investigate the scaling relations between BH mass and bulge
properties for our sample.  Bear in mind that the absolute positions
of our objects in the \mbulge\ plane are uncertain due to potential
systematic offsets in the value of $f$ and other systematic
uncertainties associated with the BH mass measurements, as well as the
unknown mass-to-light ratios of the bulges.  We will discuss how each
of these uncertainties may impact our conclusions.

\subsection{Black hole mass vs. bulge luminosity}

BH masses and absolute bulge magnitudes for our sample are shown in
Figure \ref{MBH}{\it a} as blue open circles.  We compare our
galaxies with the \mbulge\ relations published by
\cite{Bentzetal2009a} for reverberation-mapped AGNs and
\cite{Gutekinetal2009} for local inactive galaxies with dynamical BH
masses.

\cite{Bentzetal2009a} presents the \mbulge\ relation for nearby
AGNs\footnote{Bright, nearby galaxies are excluded in their analysis.} with
reverberation-based BH mass measurements (Figure \ref{MBH}; {\it a}).  
The bulge luminosities are based on GALFIT modeling of 
\emph {HST}/F550M or F547M images.  Since our virial BH masses 
are based on a radius-luminosity relation derived from these same objects, 
the BH masses are directly comparable, hopefully minimizing systematic 
errors.  However, we note that their BH masses range from 
$10^7-10^9\msun$, larger than the BH masses in our sample by
about 2 orders of magnitude.  This comparison can tell us whether
the \mbulge\ relation that is observed in massive AGNs still exists
at the low masses probed here.

In order to make this comparison, we adopt the $f$ factor from
\citet[][e.g., we boost our virial BH masses by 0.3 dex]{Onkenetal2004}.
We fit the sample with the following log-linear relation
\begin{equation}
\log(\mbh/\msun)=\alpha (M_I - M_0)+\beta,
\label{fittingrelation}
\end{equation}
where $M_0$ is the median magnitude measured from the fitted sample,
which is held constant during the fitting.  For the following fits, we
do not include galaxies with only upper limits on their bulge
magnitudes \footnote{We have tried to fit our sample including the
  upper limits using the maximum likelihood method as done in
  \cite{Gutekinetal2009}.  The answer is unchanged.}. Following
\cite{GreeneHo2006} and \cite{Tremaineetal2002}, we include the
intrinsic scatter $\epsilon_0$ in the definition of $\chi^2$ (Equation
1 of \citealt{GreeneHo2006}) so that the best fit gives a
$\chi^2=1$. Thus, we get an upper limit on the intrinsic scatter.

Compared to the Bentz et al. fit, we find a significantly steeper
slope and lower zeropoint when our sample is included.  The difference
between the two samples can be quantified by calculating the predicted
BH masses $M_{\text{predict}}$ based on our bulge luminosities and the
best-fitting relation for the Bentz et al. sample.  We calculate a
mean difference of $\langle \log(M_{\text{predict}}/\mbh) \rangle =
1.15\pm 0.03$, where the latter is the error on the mean.  
When we fit our combined samples, we find
an intrinsic scatter of $\epsilon_0=0.41-0.45$ for an assumed
scatter in virial BH masses of 0.48
\citep[e.g.,][]{Collinetal2006,Wooetal2010} to 0.4
\citep[e.g.,][]{GreeneHo2006}.  
Both calculations show that the galaxies in our sample have
systematically smaller BH masses at a fixed bulge mass compared with
the sample of Bentz et al.  Differences in stellar populations, which we
estimate would dim our bulges by $\sim 1$ mag (Paper I), are not sufficient to
explain these discrepancies.

To investigate further whether these differences represent a real
physical difference (e.g., as a function of mass), we also compare to
inactive BHs with dynamical BH masses from \cite{Gutekinetal2009}.  As
explained in Paper I, we adopt a color difference of $V-I=1.34$ mag to
shift the G\"ultekin et al. sample from the $V$-band to the $I$-band.
The average difference between the predicted and observed BH masses
based on the G\"ultekin et al. fit is
$\langle\log(M_{\text{predict}}/\mbh) \rangle=0.85\pm0.04$.  If we fit
our sample combined with that of G\"ultekin et al., the intrinsic
scatter increases significantly (see Table \ref{fittingresult}).  This
scatter is driven by a systematic shift between our sample and the
G\"ultekin et al. sample. For a given BH mass, our galaxies have
larger bulge luminosities.  This shift may be real, or it may be
explained by younger stars, and thus lower mass-to-light ratios, in
our sample galaxies (Paper I). In order to investigate the magnitude
of this uncertainty, we estimate dynamical masses $M_{\rm{dyn}}$ for a
subset of our sample, as described below. Dynamical mass should be a
good approximation of the stellar mass of the bulge $M_{\rm{bulge}}$,
since in general stars should dominate the gravitational potential on
the scale of the bulge.
 
 \subsection{Dynamical masses of the bulges}

Dynamical masses for the bulges are estimated as $M_{\rm{dyn}}=5
r_e\sigma^2/G$, where $r_e$ is the effective radius and $\sigma$ is
the velocity dispersion of the bulge (Paper I). Here we use a
prefactor of $5$ to be consistent with \cite{Cappellarietal2006} and
\cite{Sanietal2011}. Dynamical masses for the sample of G\"ultekin et al.
are from \cite{HaringRix2004}.  For the pseudobulges from
\cite{Kormendyetal2011}, we use effective radii from
\cite{Sanietal2011}. We also include $6$ additional pseudobulges
identified by \cite[][Figure \ref{MBH} {\it b}]{Sanietal2011}.  For the
galaxies with stellar velocity dispersion measurements, the average
difference between the predicted BH masses (based on the best fitting 
relation of G\"ultekin et al.'s sample and estimated dynamical masses) and
observed values is $\langle \log(M_{\text{predict}}/ \mbh)
\rangle=0.79\pm0.09$. This offset is significantly larger than any
reasonable formal uncertainty in the virial BH masses ($0.4-0.48$
dex).  Furthermore, even after correcting for stellar populations,
our sample follows a steeper relation between $M_{\rm BH}$ and
$M_{\rm bulge}$ than the inactive ellipticals taken alone (Table
\ref{fittingrelation}). This is consistent with our observed
Faber-Jackson relation in Paper I (Figure 9), in that at a fixed
bulge magnitude, pesudobulges have smaller stellar velocity
dispersions. 

When a dark matter halo is included, the prefactor in the formula 
for dynamical mass decreases by $\sim12\%$ \citep[][]{Cappellarietal2006}.
Furthermore, \cite{Tayloretal2010} show that the normalization of dynamical 
mass formula for bulges with S\'ersic index $n=2$ (as is the case for our 
sample of predominantly pseudobulges) is be larger by $\sim40\%$. 
Combining the above two effects 
would increase our estimated dynamical mass by $\sim28\%$. Considering 
the large uncertainty in the virial BH masses (a factor of 3), this is 
unlikely to change our results. Furthermore, 
if we were to boost the dynamical masses of the galaxies in our sample, 
 the difference between our sample and G\"ultekin et al.'s 
sample would be larger. 

Interestingly, the galaxies without extended disks show the largest
offset (red triangles in Figure \ref{MBH}). They are also the largest
outliers in the Faber-Jackson relation in Paper I.  Rather than
scaling like pseudobulges, these galaxies scale like spheroidals 
in the fundamental plane.  We
have no complete explanation for why these galaxies are the largest
outliers, but it is intriguing\footnote{We have also divided our
  sample into galaxies with and without a bar. They do not
  show any significant difference.}.

Finally, we look at pseudobulges (both active and inactive) and use
the Kendall's $\tau$ rank correlation coefficient to test whether
pseudobulge luminosity correlates with BH mass. The
\cite{Kormendyetal2011} sample of pseudobulges taken alone has
$\tau=0.44$ and $P=0.1$.  There is no significant \mgal\ correlation
among these galaxies.  Including our sample yields $\tau=0.13$ and
$P=0.03$. In this case there is marginal evidence for a very weak
correlation.  Thus we support the conclusion of
\cite{Kormendyetal2011} that pseudobulges show no correlation between
BH mass and bulge mass.

\section{Discussion and Conclusion}

We investigate the \mbulge\ and \mgal\ relations focused on $10^5
<\mbh/\msun< 10^6$.  Both the \mbulge\ and \mgal\ relations present
the same conclusion that these galaxies appear to have smaller BH
masses at a given bulge mass than more massive, bulge-dominated
systems.  Here we discuss the possible origins of this difference.  It
cannot be easily explained by errors in the BH mass scale. We would
have to boost the BH masses by $\sim 1$ dex, which we consider
unlikely given that the sample appears to obey an \msigma\ relation
\citep[][]{Barthetal2005, GreeneHo2006,Xiaoetal2011}.  Likewise, even
correcting for young stellar populations in these bulges, we still find that our
BHs are not consistent with the low mass extrapolation of the \mbulge\
relation for more massive BHs (consistent with
\citealt{Hu2008,Greenetal2008,GadottiKauffmann2009,Greenetal2010}).

As justified in Paper I, most of our galaxies are consistent with the
properties of pseudobulges \citep[e.g.,][]{KormendyKennicutt2004,
  FisherDrory2010}.  We believe that pseudobulges are built by secular
processes driven by disk instabilites.  If the violent merging that
builds classical bulges also establishes BH-bulge scaling relations, we
would expect different (or nonexistent) scalings for pseudobulges.  We
confirm the conclusion of \cite{Kormendyetal2011} that there is
effectively no correlation between BH mass and pseudobulge luminosity
based on our much larger sample. Furthermore, we extend this
conclusion to bulge dynamical mass.  This finding suggests that
indeed, pseudobulges have quite different properties compared with
classical bulges over a wide dynamic range, from $\sim10^5\msun$
to $\sim10^8\msun$ BHs.  However, our sample is different from the
sample of Kormendy et al. in that there is no systematic offset
between the Kormendy et al. and the G\"ultekin et al. sample.  Thus we
see additional tentative evidence for larger systematic
differences as a function of mass.

Taking our sample alone, the host galaxies have a wide range of bulge
magnitudes over a very limited range in BH mass. In other words, the
\mbulge\ relation flattens out at the low-mass end.  This seems to be
consistent with the results from simulations of
\cite{VolonteriNatarajan2009} for massive BH seed models. If BH seeds
can be as massive as $\sim10^5\msun$, the low-mass BHs in our sample
may be remnants of BH seeds that have not evolved much 
since they were formed. Consistent with our previous claims, we suggest that 
secular processes are relatively inefficient at feeding central BHs. 
Thus bulges that are built secularly maintain a low BH mass independent of 
the growth of their surrounding bulge.

\section*{Acknowledgements}
Y.-F.J thanks Chien Peng for help in using GALFIT and Minjin Kim for
helpful discussions on fitting the images. We also thank the anonymous 
referee for valuable comments to improve the manuscript. This work was supported by 
the Carnegie Institution for Science and by NASA 
grant HST-GO-11130.01 from the Space Telescope Science Institute, which
is operated by AURA, Inc., under NASA contract NAS5-26555.

\end{document}